\documentclass[3p,times,twocolumn]{elsarticle}

\usepackage{ecrc}

\volume{00}

\firstpage{1}

\journalname{Nuclear Physics B Proceedings Supplement}

\runauth{M. Bonvini et al.}

\jid{nuphbp}

\jnltitlelogo{DESY 13-005}

\usepackage{amssymb}
\usepackage{amsmath}
\usepackage{booktabs}
\usepackage{xcolor}
\usepackage[pdftex]{hyperref}
\allowdisplaybreaks

\usepackage[figuresright]{rotating}

\newcommand{\mus}{\mu_s}
\newcommand{\as}{\alpha_s}

\newcommand{\Ord}{\mathcal{O}}
\newcommand{\Lum}{\mathcal{L}}
\newcommand{\Cscet}{C_{\rm\scriptscriptstyle SCET}}
\newcommand{\Cqcd}{C_{\rm\scriptscriptstyle QCD}}
\newcommand{\sigscet}{\sigma_{\rm\scriptscriptstyle SCET}}
\newcommand{\sigqcd}{\sigma_{\rm\scriptscriptstyle QCD}}
\newcommand{\Sud}{\mathcal{S}}
\def\beq{\begin{equation}}
\def\eeq{\end{equation}}
\def\({\left(}
\def\){\right)}
\def\[{\left[}
\def\]{\right]}

\begin{document}

\begin{frontmatter}



\dochead{}

\title{The scale of soft resummation in  SCET vs perturbative QCD}


\author[desy]{Marco Bonvini}
\author[milano]{Stefano Forte\corref{speaker}}
\author[roma1]{Margherita Ghezzi}
\author[genova]{Giovanni Ridolfi}
\address[desy]{Deutsches Elektronen-Synchroton, DESY, Notkestra{\ss}e 85, D-22603 Hamburg, Germany}
\address[milano]{Dipartimento di Fisica, Universit\`a di Milano and INFN, Sezione di Milano, Via Celoria 16, I-20133 Milano, Italy}
\address[roma1]{Dipartimento di Fisica, Sapienza Universit\`a di Roma and INFN, Sezione di Roma, Piazzale Aldo Moro 2, I-00185 Roma, Italy}
\address[genova]{Dipartimento di Fisica, Universit\`a di Genova and INFN, Sezione di Genova, Via Dodecaneso 33, I-16146 Genova, Italy}
\cortext[speaker]{Speaker.}

\begin{abstract}
We summarize and extend previous results on
the comparison of threshold resummation, performed, using
soft-collinear effective theory (SCET), in the Becher-Neubert approach,
to the standard perturbative QCD formalism based on factorization and
resummation of Mellin moments of partonic cross sections.
We show that the logarithmic accuracy of this SCET 
result can be extended by half a logarithmic order, thereby bringing it
in full agreement with the standard QCD result if a suitable choice
is made for the soft scale $\mu_s$ which characterizes the SCET
result.
We provide a master formula relating the two approaches for other scale choices.  
We then show that with the Becher-Neubert scale choice the Landau
pole, which in the perturbative QCD approach is usually removed
through power- or exponentially suppressed terms, in the SCET approach
is removed by logarithmically subleading terms which break
factorization. Such terms  may become leading for generic choices of
parton distributions, and are always leading when resummation is used
far enough from the hadronic threshold.
\end{abstract}


\end{frontmatter}


\section{Soft resummation and scale choices}
\label{sec:intro}
Threshold resummation~\cite{Sterman:1986aj,Catani:1989ne} plays an important role in extending and
stabilizing the accuracy of perturbative results, and it may be of
some  relevance even for hadronic processes which are quite far 
from threshold~\cite{Bonvini:2010tp,Bonvini:2012an}, 
due to the fact that the underlying partonic process can be
rather closer to threshold than the hadronic one~\cite{Catani:2001ic}. All-order resummed
results are known to lead to a divergent series when expanded out in
powers of the strong coupling: this is physically due to the fact that
resummation is obtained by choosing as a scale of the parton-level
process the maximum energy of the radiated partons~\cite{Contopanagos:1996nh,Forte:2002ni}, which tends to zero in
the threshold limit. 
The divergence can be tamed by introducing  suitable subleading
contributions, such as exponentially suppressed terms outside the
physical kinematic region~\cite{Catani:1996yz}, or power-suppressed
terms~\cite{Forte:2006mi,Abbate:2007qv}.  

In Ref.~\cite{Becher:2006nr}
it was suggested, within the context of a SCET approach to threshold
resummation, that the divergence can be tamed by making a hadronic
choice of scale. In SCET this is possible because resummed results are
characterized by a ``soft scale'' $\mu_s$: the Becher-Neubert (BN)
scale choice consists of expressing $\mu_s$ in terms of kinematic
variables of the hadronic scattering process. The meaning of this
choice is not obvious in
the conventional QCD approach, where, because of perturbative
factorization, the partonic cross section, which is being resummed, is
independent of the hadronic kinematic variables. 

\begin{table*}[t]
\begin{center}
\begin{tabular}{lcccr}
  QCD:  & $A(\as)$ & $D(\as)$ & $\bar g_0(\as)$ & accuracy: $\as^n \ln^kN$\\
  SCET: & $\Gamma_{\rm cusp}(\as)$ & $\gamma_W(\as)$ & $H$, $\tilde s_{\rm DY}$
  & accuracy: $\as^n \ln^k(\mus/M)$\\
  \midrule
  LL     & 1-loop & --- & tree-level & $k= 2n$ \\
\addlinespace[0.8\defaultaddspace]
  NLL*   & 2-loop & 1-loop & tree-level & $2n-1\le k\le 2n$ \\
  NLL    & 2-loop & 1-loop & 1-loop & $2n-2\le k\le 2n$ \\
\addlinespace[0.8\defaultaddspace]
  NNLL*  & 3-loop & 2-loop & 1-loop & $2n-3\le k\le 2n$ \\
  NNLL   & 3-loop & 2-loop & 2-loop & $2n-4\le k\le 2n$
\end{tabular}
\caption{Orders of logarithmic approximations and accuracy of the
  predicted logarithms in perturbative QCD (first header) and SCET
  (second header). The last columns refers to the content of the
  coefficient function.
  In Ref.~\cite{Becher:2007ty}, only the N$^k$LL* counting is considered for the SCET resummation.}
\label{tab:count}
\end{center}
\end{table*}

In Ref.~\cite{Bonvini:2012az} we have clarified this issue by
explicitly exhibiting a relation between the $\mu_s$ dependent
resummed SCET result, and the standard ($\mu_s$ independent) QCD
expression.
Specializing to the BN scale choice (while taking the
Drell-Yan process~\cite{Becher:2007ty} as an example) 
we were able to show that in the SCET
result, with the BN scale choice, the divergence is removed through terms which are
logarithmically subleading in comparison to the logarithmic accuracy of
the SCET result of Ref.~\cite{Becher:2007ty}, which is by half a
logarithmic order  lower
than that of the standard QCD result. 

Here we show that the accuracy of the SCET result of Ref.~\cite{Becher:2007ty} can actually be
increased to the same level as that of the QCD result, and we rederive,
within this higher accuracy, our master formula, which thus becomes
particularly transparent. We then use this improved master formula to
discuss various problems related to the BN scale choice.

\section{Resummation of the Drell-Yan process in SCET and QCD}

We consider for definiteness inclusive Drell-Yan  production, but
the same discussion 
applies to other processes, such as Higgs production in gluon-gluon
fusion, with minimal modifications.
The dimensionless invariant mass distribution
$\sigma(\tau,M^2)=\frac{1}{\tau \sigma_0} \frac{d\sigma_{\rm DY}}{dM^2}$,
with $M$ the invariant mass of the pair and $\sigma_0$ the leading order
partonic cross section, can be written schematically
(omitting a sum over partons) in factorized form as
\beq\label{sigma}
\sigma(\tau,M^2)=\int_\tau^1\frac{dz}{z}\,\Lum\(\frac{\tau}{z}\)\,C(z,M^2),
\eeq
where $\Lum$ is the parton luminosity, $s$ the hadronic 
center-of-mass energy squared, and $\tau=\frac{M^2}{s}$,
so that the
hadronic  threshold limit is $\tau\to1$. The perturbatively
computable coefficient function
$C(z,M^2)$ is 
normalized so that $C(z,M^2)=\delta(1-z)$
at leading order in the strong coupling
$\as$.
In the sequel, without significant loss of generality,  
we shall always choose the renormalization and
factorization scales $\mu^2_F=\mu^2_R=M^2$.

Standard QCD resummation follows from a Mellin-space renormalization-group
argument~\cite{Contopanagos:1996nh,Forte:2002ni}: indeed, at the
resummed level, both the convolution Eq.~\eqref{sigma} and the gluon
radiation phase space factorize, so at the resummed level
one may write
\beq\label{melsigma}
\sigma(N,M^2)=\int_0^1d\tau\,\tau^{N-1}\sigma(\tau,M^2) = \Lum(N)\, C(N,M^2)
\eeq
where 
the $N$-space resummed coefficient function has the form~\cite{Sterman:1986aj,Catani:1989ne}
\beq\label{Cqcd}
\Cqcd(N,M^2)=\bar g_0\(\as(M^2)\) \exp\bar\Sud\left(M^2,\frac{M^2}{N^2}\right)
\eeq
where
\begin{align}
&\bar\Sud\left(M^2,\frac{M^2}{N^2}\right)
=\int_0^1dz\,\frac{z^{N-1}-1}{1-z} \label{sqcddef} \\
&\times\Bigg[
\int_{M^2}^{M^2(1-z)^2} \frac{d\mu^2}{\mu^2}
 2A\(\as(\mu^2)\) + D\(\as([1-z]^2M^2)\) \Bigg].\nonumber
\end{align}
The functions $\bar g_0(\as)$, $A(\as)$ and $D(\as)$ are 
power series in $\as$, with $\bar g_0(0)=1$ and $A(0)=D(0)=0$. 

Because
resummation is obtained through exponentiation, it might seem natural to also
exponentiate the function $\bar g_0$. However, unlike
$\bar\Sud(M^2,M^2/N^2)$, $\bar g_0$ is independent of
$N$ and only depends on $\as(M^2)$. As a consequence, it turns out
that simply including an extra term in $\bar g_0$ at each order increases
the logarithmic accuracy of the coefficient function by half a
logarithmic order. This is summarized in Table~\ref{tab:count}, where
the logarithmic accuracy obtained by including a given number of terms
in $\bar g_0$, $A$, and $D$ is summarized. A given accuracy means that all
and only
the logarithmically
enhanced contributions to the coefficient function  listed in the last
column are correctly predicted. At leading logarithmic (LL)
accuracy only the largest power of $\ln N$ at each order in $\as$ is predicted;
adding one order in each of the functions $A$, $D$, and $\bar g_0$
one then obtains the next-to-leading logarithmic (NLL) accuracy, which correctly
predicts two powers more, and so on to N$^k$LL accuracy. 
However, if $\bar g_0$ is exponentiated and a power counting is performed
at the level of exponents, it may seem more natural to include  one
less order in $\bar g_0$. This results in the  N$^k$LL*  accuracy, also
shown in table, which is lower
by one power of $\ln N$
at each order in $\as$ than the N$^k$LL accuracy.

The resummed SCET expression  for Drell-Yan pair production 
is given by~\cite{Becher:2007ty}%
\footnote{The resummed expression as given in Ref.~\cite{Becher:2007ty}
actually depends on several hard energy scales,
which here for simplicity are all taken to be equal to the hard scale $M^2$.}
\beq\label{Cscet}
\Cscet(z,M^2,\mus^2) = H(M^2)U(M^2, \mus^2) S(z, M^2,\mus^2)
\eeq
where $H(M^2)$ (hard function) is a power series in
$\as(M^2)$,
\beq\label{eq:soft_func}
S(z,M^2, \mus^2) = 
\tilde s_{\rm DY}\(\ln\frac{M^2}{\mus^2}+\frac{\partial}{\partial\eta},\mus\)
\frac{(1-z)^{2\eta}}{1-z}
\frac{e^{-2\gamma \eta}}{\Gamma(2\eta)}
\eeq
(soft function) depends on
\beq\label{etadef}
\eta=\int_{M^2}^{\mus^2}\frac{d\mu^2}{\mu^2}\,\Gamma_{\rm cusp}\(\as(\mu^2)\),
\qquad \Gamma_{\rm cusp}(\as)=A(\as),
\eeq
$\tilde s_{\rm DY}(L,\mus)$ is a series in $\as(\mus^2)$ with
$L$-dependent coefficients,  and
\begin{multline}\label{udef}
\ln U(M^2,\mus^2) = -\int_{M^2}^{\mus^2} \frac{d\mu^2}{\mu^2}\\
   \times \Bigg[\Gamma_{\rm cusp}\(\as(\mu^2)\) \ln\frac{\mu^2}{M^2}
    - \gamma_W\(\as(\mu^2)\) \Bigg],
\end{multline}
where $\gamma_W(\as)$ is also
a series  in $\as$, with $\gamma_W(0)=0$.  The scale $\mus$ is
a soft matching scale of the effective theory, and $\Cscet$ formally
does not depend on it, up to subleading terms.  However, the SCET
result resums powers of $\ln\frac{\mus}{M}$, so this choice of scale
determines what  is being resummed.  

Again, a given logarithmic accuracy is obtained by including a
finite number of terms in the perturbative expansion of the functions which determine the
resummed result, namely $\Gamma_{\rm cusp}$, $\gamma_W$, $H$ and
$\tilde s_{\rm DY}$, according to Table~\ref{tab:count}.
In Ref.~\cite{Becher:2007ty}, only the  N$^k$LL* accuracy was
considered: in fact,  the order called N$^k$LL* in
Tab.~\ref{tab:count} is actually referred to 
N$^k$LL Ref.~\cite{Becher:2007ty}, which might be the source of some
confusion.  
Computations using either of these two
definitions of the logarithmic accuracy have been presented in the past,
either in the contex of QCD (see e.g.\ Ref~\cite{Dasgupta:2003iq},
where N$^k$LL* is referred to as N$^k$LL$_{\ln R}$) or SCET (see
e.g.\ Ref.~\cite{NNLL'}, where N$^k$LL is referred to as  N$^k$LL$^\prime$).

Here
we show that in fact the N$^k$LL* of Ref.~\cite{Becher:2007ty} can be
promoted to 
higher  N$^k$LL accuracy, by
inclusion of the terms listed in the table. This result is
obtained in the next Section, by explicitly computing the relation
between this improved version of the SCET result, and the QCD result.

\section{Comparison at NNLL}
\label{sec:comparison}

An analytic comparison between the QCD and SCET
resummation formalisms can be performed~\cite{Bonvini:2012az}
in $N$ space, where the QCD result is naturally constructed, and where
it admits a convergent perturbative expansion in powers of
$\as$. Namely, we determine the ratio $C_r(N,M^2,\mu^2_s)$
between the QCD and SCET expressions, Eqs.~\eqref{Cqcd} and \eqref{Cscet}:
\beq\label{CCC}
\Cqcd(N,M^2) = C_r(N,M^2,\mu^2_s)\, \Cscet(N,M^2,\mus^2).
\eeq
In Ref.~\cite{Bonvini:2012az} we computed $C_r(N,M^2,\mu^2_s)$ to
NNLL, using the definition of NNLL of Ref.~\cite{Becher:2007ty}, which
in Tab.~\ref{tab:count} we call NNLL*. Here we show that the
accuracy of the SCET expression can be upgraded, and the comparison
can be carried out at full  NNLL according to the definition of
Table~\ref{tab:count}.

We first rewrite the QCD result in the more
convenient form~\cite{Bonvini:2012az}
\beq\label{Cqcd2}
\Cqcd(N,M^2)=\hat g_0\(\as(M^2)\) \exp\hat\Sud_{\rm\scriptscriptstyle QCD}\left(M^2,\frac{M^2}{\bar N^2}\right)
\eeq
with $\bar N = N e^\gamma$ and
\begin{align}
&\hat g_0(\as)=\bar g_0(\as) \exp\[2\zeta_2A(\as)+\frac83\zeta_3\beta_0\frac{C_F}\pi\as^2\],
\label{ghatdef}\\
&\hat{\Sud}_{\rm\scriptscriptstyle QCD}\(M^2,\frac{M^2}{\bar{N}^2}\)
=\int_{M^2}^{M^2/\bar{N}^2}\frac{d\mu^2}{\mu^2}\,
\Bigg[A\(\as(\mu^2)\)\ln\frac{M^2}{\mu^2\bar{N}^2}\nonumber\\
&\qquad\qquad\qquad\qquad\qquad\qquad\quad+\hat D\(\as(\mu^2)\)\Bigg],
\label{shatdef}\\
&A(\as)=\frac{A_1}{4}\as+\frac{A_2}{16}\as^2+\frac{A_3}{64}\as^3+\Ord(\as^4),\label{Aexp}\\
&\hat D(\as) = \frac12 D(\as) - 2\zeta_2 \frac{C_F}{\pi}\beta_0 \as^2
=\hat D_2\as^2+\Ord(\as^3),
\end{align}
and $\beta_0=(11C_A-2n_f)/(12\pi)$.

The functions $A$, $D$ and $\bar g_0$ are computed at NNLL order
according to Table~\ref{tab:count}; the relevant coefficients can be found in 
Ref.~\cite{Bonvini:2012az},
except the two-loop contribution to $\bar g_0$, which can be determined
by matching the expansion of Eq.~\eqref{Cqcd} to the NNLO Drell-Yan
cross section in Ref.~\cite{Hamberg:1990np},
\begin{align}
\bar g_0(\as) &= 1 + \frac{\as}\pi C_F\(2 \zeta_2 - 4\) \nonumber\\
& + \frac{\as^2}{\pi^2} \frac{C_F}{16}
\Bigg[C_A \(-\frac{12}{5} \zeta_2^2 + \frac{592}{9} \zeta_2 + 28 \zeta_3 - \frac{1535}{12}\)\nonumber\\
&\qquad\qquad + C_F \(\frac{72}{5} \zeta_2^2 - 70 \zeta_2 - 60 \zeta_3 + \frac{511}{4}\) \nonumber\\
&\qquad\qquad + n_f \(8 \zeta_3 - \frac{112}{9} \zeta_2 + \frac{127}{6}\)\Bigg],
\end{align}
and the coefficient $A_2$, which we give here for completeness
\beq
A_2=\frac{4C_F}{\pi^2}\[\(\frac{67}{9}-2\zeta_2\)C_A-\frac{10}{9}n_f\].
\eeq
The explicit expression of $A_3$ is not needed here.

On the other hand, at NNLL the Mellin transform of the SCET result, Eq.~\eqref{Cscet},
can be written as~\cite{Bonvini:2012az}
\begin{multline}
\Cscet(N,M^2,\mus^2)= \hat H(M^2) \, E\(\frac{M^2}{\bar N^2},\mus^2\)\\
\times\exp\hat\Sud_{\rm\scriptscriptstyle SCET}(N,M^2,\mus^2),
\label{Cscetnew}
\end{multline}
with
\begin{align}
&\hat H(M^2) = H(M^2) \exp\[\frac{\zeta_2}{2} \frac{C_F}{\pi} \as(M^2) \],\\
&E\(\frac{M^2}{\bar N^2},\mus^2\) = \tilde s_{\rm DY}\(\ln\frac{M^2}{\mus^2\bar N^2},\mus^2\)
\exp\[-\frac{\zeta_2}{2} \frac{C_F}{\pi} \as(\mus^2)\],\label{eq:Edef}\\
&\hat\Sud_{\rm\scriptscriptstyle SCET}(N, M^2,\mus^2)=
\int_{M^2}^{\mus^2} \frac{d\mu^2}{\mu^2}
\Bigg[\Gamma_{\rm cusp}\(\as(\mu^2)\) \ln\frac{M^2}{\mu^2\bar{N}^2} \nonumber\\
&\qquad\qquad\qquad\qquad\qquad\qquad+\hat\gamma_W\(\as(\mu^2)\) \Bigg],
\label{scetsfun}\\
&\hat \gamma_W(\as) = \gamma_W(\as) - \frac{\zeta_2}{2} \frac{C_F}{\pi}\beta_0\as^2,
\end{align}
and $\hat \gamma_W(\as) = \hat D(\as)$ at this order.

In comparison to Ref.~\cite{Bonvini:2012az}, we now also include
the two-loop contributions to the functions $H$ and $\tilde s_{\rm DY}$,
which were given explicitly in Ref.~\cite{Becher:2007ty}.
Note that, in
order to be accurate to order $\as^2$, the definition of the function
$E$ slightly differs from Ref.~\cite{Bonvini:2012az}.

Putting everything together we find
\beq\label{eq:Cr}
C_r(N,M^2,\mus^2) = \frac{\hat g_0\(\as(M^2)\)}{\hat H(M^2) E\(\frac{M^2}{\bar N^2},\mus^2\)}
\exp\hat\Sud\(\mus^2,\frac{M^2}{\bar N^2}\)
\eeq
with
\begin{multline}
\hat\Sud\(\mus^2,\frac{M^2}{\bar N^2}\)
=\int_{\mus^2}^{M^2/\bar{N}^2}\frac{d\mu^2}{\mu^2}\,
\Bigg[A\(\as(\mu^2)\)\ln\frac{M^2}{\mu^2\bar{N}^2}\\
+\hat D\(\as(\mu^2)\)\Bigg].
\label{Shat}
\end{multline}
It is easy to see that
\beq\label{wcrat}
\frac{\hat g_0\(\as(M^2)\)}{\hat H(M^2) E(M^2,M^2)} = 1+\Ord\(\as^3(M^2)\),
\eeq
so to NNLL accuracy Eq.~\eqref{eq:Cr} can be written
\beq\label{eq:Cr2}
C_r(N,M^2,\mus^2) = \frac{E(M^2,M^2)}{E\(\frac{M^2}{\bar N^2},\mus^2\)}
\exp\hat\Sud\(\mus^2,\frac{M^2}{\bar N^2}\).
\eeq
Using the 2-loop
expression of $\tilde s_{\rm DY}$ from Ref.~\cite{Becher:2007ty}
in Eq.~\eqref{eq:Edef}, we find
\beq
E\(\frac{M^2}{\bar N^2},\mus^2\)=1+E_1(L)\as(\mus^2)+E_2(L)
\as^2(\mus^2)+\Ord(\as^3)
\eeq
where
\begin{align}
E_1(L)&=\frac{A_1}{8}L^2, \\
E_2(L)&=\frac{A_1^2}{128}L^4  
-\frac{L^3}{3}\frac{A_1}{8}\beta_0
+\frac{L^2}{2}\frac{A_2}{16}
+L\hat D_2
\nonumber\\
&+\frac{C_AC_F}{\pi^2}\Bigg[\frac{607}{324}
+\frac{67}{144}  \zeta_2-\frac{3}{4}  \zeta_2^2
-\frac{11}{72}  \zeta_3\Bigg]
\nonumber\\
&+ \frac{C_Fn_f}{\pi^2} \[-\frac{41}{162}
-\frac{5}{72} \zeta_2+\frac{\zeta_3}{36}
\],
\label{eq:Eser}
\end{align}
and
\beq
L \equiv \ln\frac{M^2}{\mus^2\bar N^2}.
\eeq
Note that $L=0$ when the two arguments of $E$ are equal to each other.

Eq.~\eqref{eq:Cr2} establishes our first new result. Indeed,
it is immediate to check that 
$C_r(N,M^2,\mus^2)=1$ for $\mus=M/\bar N$, up to subleading (NNNLL*)
terms. 
This means that with this
scale choice the SCET
result now reproduces the QCD result to full NNLL accuracy, rather
than to the lower NNLL* accuracy of Ref.~\cite{Becher:2007ty}.

We are however interested in studying $C_r$ for generic scale choices,
and in particular with the BN scale choice.
The result becomes especially transparent by casting the ratio
$E(M^2,M^2)/E\(\frac{M^2}{\bar N^2},\mus^2\)$ Eq.~\eqref{eq:Cr2}
in the form of an exponential of an integral, of the same kind as the
form adopted in
Eq.~\eqref{Shat}. This can be done at the price of including
terms of order $\as^3$ or higher in the ratio,
which is allowed at NNLL. The ensuing expression of $C_r$ is
particularly simple and suitable for analytic comparisons. It should
however be kept in mind that
a numerical  comparison of the SCET and QCD expressions 
should rather be performed using
the exact expression Eq.~\eqref{eq:Cr2}, and possibly also retaining
the subleading terms in Eq.~(\ref{wcrat}).

We get
\begin{align}
&\ln\frac{E(M^2,M^2)}{E\(\frac{M^2}{\bar N^2},\mus^2\)}
=-\as(\mus^2)\frac{A_1}{4}\frac{L^2}{2}
\nonumber\\\label{eexp}
&\qquad
+\as^2(\mus^2)\[\beta_0\frac{A_1}{8}\frac{L^3}{3}-\frac{A_2}{16}\frac{L^2}{2}
-\hat D_2L\]+\Ord(\as^3).
\end{align}
Using
\beq
\frac{L^{k+1}}{k+1}=\int_{\mus^2}^{M^2/\bar N^2}\frac{d\mu^2}{\mu^2}\,
\ln^k\frac{M^2}{\mu^2\bar N^2}
\eeq
and taking the running of $\as$ into account, we finally obtain
\begin{multline}
C_r(N,M^2,\mus^2)=\exp\int_{\mu_s^2}^{M^2/\bar{N}^2}
\frac{d\mu^2}{\mu^2}\,\ln\frac{M^2}{\mu^2\bar{N}^2}
\\
\times\[A\(\as(\mu^2)\)-\frac{A_1\as(\mu^2)}{4}-\frac{A_2\as^2(\mu^2)}{16}\],
\label{Srexp}
\end{multline}
which is our NNLL master QCD-SCET comparison formula. It generalizes
to full NNLL the result of Ref.~\cite{Bonvini:2012az}. Its most
notable feature, which determines the relative accuracy of the
comparison, is that (recall the expansion Eq.~\eqref{Aexp}) 
the exponent in Eq.~(\ref{Srexp}) is 
of order $\as^3$. Note that this is however due to the exponentiation
Eq.~(\ref{eexp}). If one does not exponentiate (as in the original
SCET expression), when expanding $C_r$ in powers of $\as$,
terms proportional to $A_1$ and $A_2$ only
cancel up to $\Ord(\as^2)$.

\section{The Becher-Neubert scale choice}

As briefly discussed in Sect.~\ref{sec:intro}, the BN approach is
based on the idea of choosing for $\mu_s$ a scale determined by
hadronic, rather than partonic kinematics, namely
$\mus = M(1-\tau)$. In Ref.~\cite{Becher:2006nr,Becher:2007ty} a more
general choice $\mus = M(1-\tau)g(\tau)$ was considered, with
$g(\tau)={\rm const.}+\Ord(1-\tau)$: the distinction may be relevant
for phenomenology, but it is immaterial for our present goal, which is
to determine the logarithmic accuracy of the SCET result with this
scale choice. 

Since the variable $\tau$ refers to hadron
kinematics, the comparison can only
be performed at the level of the physical cross section, Eq.~\eqref{sigma}.
We therefore define
\begin{align}
&\sigqcd(\tau,M^2)=\int_\tau^1\frac{dz}{z}\,\Lum\(\frac{\tau}{z}\)\,\Cqcd(z,M^2),
\label{sigmaQCD}
\\
&\sigscet(\tau,M^2)=\int_\tau^1\frac{dz}{z}\,\Lum\(\frac{\tau}{z}\)\,\Cscet(z,M^2,M^2(1-\tau)^2),
\label{sigmascet}
\end{align}
where $\Cqcd(z,M^2)$ and $\Cscet(z,M^2,\mus^2)$ are the inverse Mellin
transforms of Eqs.~\eqref{Cqcd2} and \eqref{Cscetnew}, respectively,
and in the SCET case after performing the inverse Mellin transform at fixed
$\mu_s$ we have set $\mu_s=M(1-\tau)$.
Of course, $\Cqcd(z,M^2)$  is given by a  divergent series in powers
of $\as(M^2)$, so it should be understood as the order-by-order
Mellin inversion up to arbitrarily high but finite order.
Using Eq.~\eqref{CCC} we find
\beq\label{eq:sigma_master}
\sigqcd(\tau,M^2) = \int_\tau^1 \frac{dz}z \, \sigscet\(\frac{\tau}{z},M^2\)
\,C_r(z,M^2,M^2(1-\tau)^2)
\eeq
where $C_r(z,M^2,M^2(1-\tau)^2)$ is the inverse Mellin transform of
Eq.~\eqref{eq:Cr2}, performed at fixed $\mu_s$ and evaluated at
$\mu_s=M(1-\tau)$.

In order to compute $C_r(z,M^2,M^2(1-\tau)^2)$ 
it is convenient to rewrite Eq.~\eqref{eq:Cr2} as
\beq\label{eq:Cr3}
C_r(N,M^2,\mus^2) = 
\frac{E(M^2,M^2)}{E(\mus^2,\mus^2)}
\frac{E(\mus^2,\mus^2)}{E\(\frac{M^2}{\bar N^2},\mus^2\)}
\exp\hat\Sud\(\mus^2,\frac{M^2}{\bar N^2}\).
\eeq
The first ratio is just a function of $\as(M^2)$ and $\as(\mus^2)$,
independent of $N$, while
\beq
\frac{E(\mus^2,\mus^2)}{E\(\frac{M^2}{\bar N^2},\mus^2\)}
\exp\hat\Sud\(\mus^2,\frac{M^2}{\bar N^2}\)
=1+F_r\(\as(\mus^2),L\),
\eeq
where $F_r(\as,L)$ is of order $\as^3$. 

The inverse Mellin transform can be now computed using the results of
Ref.~\cite{Bonvini:2012az}.
We find
\begin{multline}
\sigqcd(\tau,M^2) = \frac{E(M^2,M^2)}{E(\mus^2,\mus^2)}
\Bigg[\sigscet(\tau,M^2) \\
+ \left.F_r\(\as(\mus^2),2\frac{\partial}{\partial\xi}\)
\Sigma(\tau,M^2,\xi)\right|_{\xi=0}\Bigg],
\label{eq:sigma_comparison}
\end{multline}
where
\beq
\Sigma(\tau,M^2,\xi) = \frac{(1-\tau)^{-\xi}}{e^{\gamma\xi}\Gamma(\xi)}
\int_\tau^1 \frac{dz}z \, \sigscet\(\frac{\tau}{z},M^2\)\, \ln^{\xi-1}\frac1z
\eeq
for $\mus=M(1-\tau)$.
We have shown in App.~B of Ref.~\cite{Bonvini:2012az} that
$\Sigma(\tau,M^2,\xi)$
can be expressed in terms of derivatives of $\sigscet(\tau,M^2)$ with respect to
$\ln(1-\tau)$, up to terms suppressed by positive powers of $(1-\tau)$:
\beq\label{eq:Sigma_ser}
\Sigma(\tau,M^2,\xi) =
\sum_{k=0}^\infty c_k(\xi) \frac{d^k\sigscet(\tau,M^2)}{d\ln^k(1-\tau)}
\[1+\Ord(1-\tau)\],
\eeq
with coefficients $c_k$ which do not depend on $\tau$.
It follows that the term proportional to $F_r$ 
in Eq.~\eqref{eq:sigma_comparison}
does not contain any extra logarithmic enhancement with respect to
$\sigscet(\tau,M^2)$.
On the other hand
\begin{align}
\frac{E(M^2,M^2)}{E(\mus^2,\mus^2)}
&=1+E_2(0)\[\as^2(M^2)-\as^2(\mus^2)\]+\ldots
\label{EE} \\
&=1+4E_2(0)\beta_0\as^3(M^2)\ln(1-\tau)+\Ord(\as^4).
\nonumber
\end{align}

We conclude that the leading difference between the QCD and SCET
expressions is 
\begin{multline}
\sigqcd(\tau,M^2) - \sigscet(\tau,M^2)=\\ =
\sigscet(\tau,M^2) \[\as^3 4\beta_0E_2(0)\ln(1-\tau) + \dots \],
\end{multline}
where the ellipse denotes  terms which are either of relative order
$\Ord(\as^3)$, but without any
logarithmic enhancement, or  $\Ord(\as^4)$.

Because the log counting is now done at the level of hadronic cross
sections, it  is based on counting powers of $\ln(1-\tau)$. Also, because the SCET
result violates standard QCD factorization (i.e., it does not factorize
upon Mellin transformation), the difference between $\sigqcd$ and
$\sigscet$ depends on the parton luminosity (it is not universal)
through $\sigscet$ itself. A generic leading-log term in 
$\sigscet$ has the form
\beq
\sigscet\sim \as^k\ln^{2k+p}(1-\tau),
\eeq
where $\as^k\ln^{2k}(1-\tau)$ is  due to the leading log behavior of the coefficient function, and
$\ln^p(1-\tau)$ generally comes from the parton luminosity. 

If we assume that the parton luminosity does not lead to any
logarithmic enhancement, then
\beq\label{eq:sigma_diff}
\sigqcd-\sigscet\sim \as^{k+3} \ln^{2k+1}(1-\tau)=
\as^h \ln^{2h-5}(1-\tau),
\eeq
where we have set $h=k+3$. This corresponds to a 
 NNNLL* correction. It is interesting to observe that, had we used the
 exponentiated version Eq.~(\ref{Srexp}) of $C_r$, a NNNLL, rather than
 NNNLL* correction, would have been obtained.
The argument can be generalized to the case in which $C_r$ is computed
to all orders in
$\as$ rather than just to order $\as^2$.
Indeed, no leading logarithmic enhancement arises from
the factor $1+F_r(\as(\mus^2),L)$;
the only possible source of powers of $\ln(1-\tau)$
in $C_r$ is the ratio $E(M^2,M^2)/E(\mus^2,\mus^2)$. It is easy to see,
however, that all terms in the expansion Eq.~\eqref{EE}
are at most of order $\as^k(M^2)\ln^{k-2}(1-\tau)$, with $k\ge 3$.
Thus, the conclusion
Eq.~(\ref{eq:sigma_diff}) holds to all orders in $\as$.

In conclusion, we restate  three observations which were already made
in Ref.~\cite{Bonvini:2012az}.
First, we note that the BN scale choice removes the divergence of the
perturbative expansion at the cost of introducing logarithmically
suppressed non-universal terms. This is to be contrasted with the
commonly used Minimal prescription~\cite{Catani:1996yz}, which also introduces
non-universal terms (with support outside the physically accessible
kinematic region) but are more suppressed than any power, or with the
Borel prescription~\cite{Forte:2006mi,Abbate:2007qv}, which introduces
power-suppressed but universal terms.

Second, we observe that quite in general we do expect PDFs to contain
logarithmically enhanced terms. In this case the terms introduced by
the BN scale choice to tame the perturbative divergence can become
leading or even super-leading (i.e., more logarithmically enhanced than
the leading log).

Finally, we remark that threshold resummation is often useful in
situations where $\tau$ is far from threshold, but nevertheless the
partonic subprocess is close to
threshold~\cite{Catani:2001ic,Bonvini:2010tp,Bonvini:2012an}. In this
case $M(1-\tau)\sim M$, and consequently 
$C_r$ Eq.~(\ref{eq:Cr2}) is actually leading log.

The phenomenological implications of our results remain to be
investigated. They are potentially of considerable interest, given the
increasingly important role that threshold resummation, in its various
implementations, is playing for LHC phenomenology.

\section*{Acknowledgments}
We thank Frank Tackmann and Christian Bauer for stimulating discussions.
MB wishes to thank Ben Pecjak for an interesting conversation.







\end{document}